# Characterization of photoelastic materials by combined Mach-Zehnder and conoscopic interferometry: Application to tetragonal lithium tetraborate crystals

**B.G. Mytsyk** [a], **A.S. Andrushchak** [b], **D.M. Vynnyk** [b], **N.M. Demyanyshyn** [a], **Ya.P. Kost** [a], **A.V. Kityk** [c,*]

[a] *Karpenko Physico-Mechanical Institute, 5 Naukova St., Lviv 79601, Ukraine*
[b] *Institute of Telecommunications, Radioelectronics and Electronic Engineering, Lviv Polytechnic National University, 12 Bandery St., Lviv 79046, Ukraine*
[c] *Faculty of Electrical Engineering, Czestochowa University of Technology, 17 Al. Armii Krajowej, Czestochowa PL-42200, Poland*

**Abstract.** Mach-Zehnder and conoscopic interferometry are used to explore photoelastic properties of anisotropic crystal materials. In a number of cases an application of both techniques significantly improves an accuracy of piezoop- tic and photoelastic measurements. The performance of such combined approach is demonstrated on tetragonal lithium tetraborate (LTB) single crystals, as an example. Special attention is paid to methodological and metrolo- logical aspects, such as measurement accuracy and the quantitative error analysis of the resulting measurements. Performing the interferometric measurements for different geometries of piezooptic coupling the full sets of piezooptic and photoelastic tensor constants of LTB crystals have been determined. The acoustooptic efficiency, on the other hand, has been evaluated using the magnitudes of photoelastic constants derived from the piezoop- tical measurements. For the geometries with strong photoelastic coupling LTB demonstrates quite large acous- tooptic performance with figure of merit value, $M_2$, achieving $2.12 \times 10^{-15}$ s$^3$/kg. It is several times larger than that of strontium borate crystals, nowadays the best acoustooptic material in deep-ultraviolet spectral region.

## 1. Introduction

The piezo-optic effect appears to be extremely important in the field of optoelectronics and laser engineering where the search for highly efficient photoelastic and acoustooptic materials experiences a great interest over the last several decades due to the growing number of their applications as photoelastic light modulators, acoustooptic modulators and deflectors, tunable spectral filters, Q-switches, etc. [1–8]. The efficiency of the photoelastic or acoustooptic transformation is quite anisotropic even for high-symmetry crystal materials and is defined by piezooptic and elastic tensors which are specific for each material system. Having a full set of corresponding tensor constants one may choose the most appropriate crystal orientation maximizing performance of photoelastic and acoustooptic cells applicable in optoelectronic devices. Spatial analyses and optimization procedures, reported in earlier publications, see e.g. [9,10], ultimately require an accurate determination of both absolute values and signs of all non-zero piezooptic tensor constants which generally is not a trivial task in anisotropic crystal materials. Relevant approaches are based on symmetry analysis and tensor transformations. In the methodological or phenomenological aspects they are quite similar to those applicable to electrooptic crystals aimed to maximize performance of Pockels cells made from such materials [11,12].

Experimentally, the photoelastic constants could be measured by acoustooptic techniques, applying e.g. Dixon-Cohen method [13]. Such costly and time-consuming method, however, is unable to determine signs of the photoelastic constants. Interferometry techniques represent here an alternative approach. Due to high configurability interferometers have found a broad range of applications in the fields of optical metrology and visualization [14] such as aerodynamics and plasma physics [15], optical coherence tomography [16], optical path-length metrology [17] and relevant optoelectronic devices, basically as integrated phase and amplitude modulators [18]. Optical interferometry has also proved its efficiency in characterization of parametric optical effects in crystal materials, exploring their electrooptic [11,12] piezoop- tic and photoelastic properties [9,10,19–21]. Once the elastic constants are known, the whole set of independent piezooptic and/or photoelastic tensor constants can be measured for crystals of any symmetry, as was demonstrated in a number of works applying Mach-Zehnder interferometry, see e.g. [19,20]. Interferometric methods, on the other hand, usually cannot provide an accurate determination of so-called rotating piezo-optical constants. For this reason the polarization-optical technique [21] and conoscopic interferometry [22] have recently proposed to determine relevant constants more accurately. Earlier the conoscopic interferometry has been applied also to study the linear electrooptic effect in photorefractive crystal materials [23].

* Corresponding author.
  *E-mail address:* kityk@ap.univie.ac.at (A.V. Kityk).

In this paper we present a combined Mach-Zehnder and conoscopic interferometry techniques aiming to explore photoelastic properties of anisotropic crystal materials. In a number of cases parallel application of both methods significantly improves the accuracy of piezooptic and photoelastic measurements. The performance of such combined approach is demonstrated on lithium tetraborate (LTB) $Li_2B_4O_7$ single crystals, as an example.

LTB crystals, which are characterized by tetragonal symmetry 4*mm* [24] (ICDD Card Number 84–2191) may be considered as efficient engineered optical material for a number of nonlinear optical [25] and acoustooptical [26] applications. A characteristic feature of such crystals is wide spectral transparency window extending, particularly, from deep ultraviolet ($\lambda = 170$ nm) towards medium infrared region ($\lambda = 3.3$ $\mu$m). High optical radiation resistance of LTB crystals ($\sim$40 GW/cm$^2$) [25], on the other hand, extends area of their applications towards power laser technologies. Due to this feature LTB crystals have evident advantage compared to a number of widely used highly efficient electrooptical, acoustooptical and/or nonlinear optical materials, like e.g. lithium niobate ($LiNbO_3$) or tellurium dioxide ($TeO_2$) [27,28] crystals that are characterized by significantly lower optical radiation resistance, typically about one order of magnitude or more less than LTB. In combination with high mechanical strength to uniaxial compression [29], which according to our measurements exceeds 300 kgf/cm$^2$ along the principal crystallophysic axes (*X* or *Z*) or diagonal directions (*XY* or *XZ*), it makes LTB a prospective photoelastic and acoustooptical material for powerful laser operation in the UV, visible and IR spectral regions.

Piezooptic properties of LTB crystals have been studied in several works [26,30–32], however reported there piezooptic constants (POCs) are characterized by significant discrepancies. In a number of cases they considerably differ each other by absolute values, as e.g. POC $\pi_{11}$ (−2.8 Br [31,32] against −0.31 Br [26]) and also by signs, as e.g. POC $\pi_{66}$ (1.66 Br [30] or 1.1 Br [31,32] against e.g. −1.39 Br [26]), here 1 Br = 1 Brewster = $1 \cdot 10^{-12}$ m$^2$/N. Similar ambiguities are noted also for several other POCs, $\pi_{im}$. Moreover, according to [31,32] certain POCs of LTB crystals reveal also strong dependence on the light wavelength $\lambda$. For instance, POC $\pi_{12}$ changes from −0.5 Br ($\lambda = 633$ nm) to −1.7 Br ($\lambda = 442$ nm), i.e. by more than three times what is untypical for optical transparency region. Weak dispersion of POCs in the optical transparency region is predicted, particularly, by the quantum-chemical calculations [33–35], whereas their strong wavelength dependence is expected mainly in the region of fundamental absorption edge.

Revealed ambiguities and/or discrepancies evidently suggest a reexamination of photoelastic characteristics in LTB crystals. Mainly due to this reason, in contrast to previous investigations [26,30–32], the POCs are determined in the present study combining the Mach-Zehnder and conoscopic [22,36] interferometry. Special attention is paid to methodological and metrological aspects, such as measurement accuracy and the quantitative error analysis of the resulting measurements. In contrast to previous studies [19] the Mach-Zehnder interferometer has been equipped by the optical pathlength compensator. Such setup modification increases, on the one hand, an accuracy of piezooptic measurements and, on the other hand, expands applicability of the interferometric method for samples with weak piezooptic effect, particularly, when relevant half-wave stress values exceed their mechanical strength and cannot be reached experimentally. To improve further the accuracy and provide reliability of piezooptic measurements most POCs and/or combinations of their sums have been determined employing a series of independent experimental geometries. Using the full tensor sets of POCs, $\pi_{im}$, extracted from piezooptic measurements, and elastic tensor constants, $C_{mk}$, we have determined a full tensor set of photoelastic constants, $p_{ik}$, and subsequently evaluated relevant acoustooptic efficiency (figure of merit $M_2$) of LTB crystals.

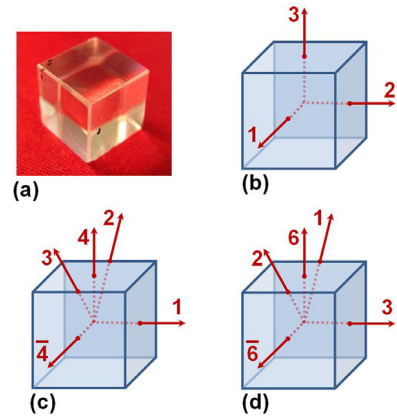

**Fig. 1.** Precisely polished sample of LTB crystal (typical view) used for piezooptic measurements [Section (a)]. Sections (b)-(d) sketch a minimal set of crystal cuts required for determination of full tensor matrices of piezooptic and photoelastic coefficients. Section (b): Crystal sample with faces cut perpendicular to the principal crystallographic axes ($X = 1$, $Y = 2$, $Z = 3$), principal-cut-sample. Section (c): $X/45°$-cut sample. Section (d): $Z/45°$-cut sample.

## 2. Experimental

Tetragonal crystals of 4*mm* symmetry are characterized by the POCs tensor having 7 independent non-zero components. For their determination three samples of different crystallographic orientations in a shape of cubes with edge length of about 7.5 mm, see Fig. 1(a), have been cut from high quality LTB single crystals and subsequently precisely polished providing high plane parallelity (better than 1′) of opposite faces in order to diminish considerably the errors caused by sample thickness nonuniformities, as described in details in Ref. [37]. Tensor analysis shows that an accurate determination of full POCs tensor matrix requires samples of three different crystallographic orientations. The most convenient ones: (i) the cubic cut with faces oriented perpendicularly to the principal crystallographic axes (hereafter the principal-cut-sample) [Fig. 1(b)]; (ii) the cubic $X/45°$-cut [Fig. 1(c)] and (iii) the cubic $Z/45°$-cut [Fig. 1(d)] have been chosen for interferometric piezooptic measurements.

Interferometry is a frequently used experimental technique for POCs measurements. Our experimental setup is based on Mach-Zehnder interferometer, see Fig. 2(a). Usually piezooptic interferometric measurements are executed at a ramping uniaxial compression, $\sigma$, applied to the sample *S* aiming to derive stress dependent light intensity, $I(\sigma)$, of the interfering beams registered by the photodetector *PD* at the interferometer output. Relevant interference fringes shift thereby in a way controlled by the uniaxial compression thus the half-wave stress value, $\sigma_{im}^{\lambda/2}$, is defined as difference of the stress magnitudes corresponding to neighbor minima and maxima of the measured $I(\sigma)$-dependence. Here indices *i* and *m* correspond to the directions of light polarization and uniaxial compression, respectively. Such methodology has been used in a number of previous studies, see e.g. [19,21]. It is applicable, however, when mechanical strength of the samples is higher than $\sigma_{im}^{\lambda/2}$-value which is not always the case. In order to remove this limitation and improve the accuracy of our measurements the optical pathlength compensator *C*, which represents a rotating parallel-plane amorphous quartz plate, has been set into one of the interferometer arms as depicted in Fig. 2(a). The optical pathlength, $\delta\Delta_k$, induced by the applied stress, $\sigma_m$, can be compensated by rotating the quartz plate by angle $\alpha$ ($\delta\Delta_k = \delta\Delta(\alpha)$) in a way to keep the output light intensity at the same level. Following the route given in [38] it is easily to show that the angular changes of the optical pathlength are defined as $\delta\Delta(\alpha) = d[1 - \cos\alpha - n + (n^2 - \sin^2\alpha)^{1/2}]$, where $n = 1.45702$, $d = 0.3$ mm are the refractive index and thickness of the quartz plate, respectively. Accordingly, $\delta\Delta(\alpha)$-dependence was calculated, verified within the calibrated procedure and tabulated. Tak-ing into account that the half-wave stress value $\sigma_{im}^{\lambda/2}$ corresponds to the stress induced optical pathlength of $\lambda/2$, it reads then as $\sigma_{im}^{\lambda/2} = \sigma_m \lambda/(2\delta\Delta(\alpha))$, where $\sigma_m$ is the stress applied to the sample.

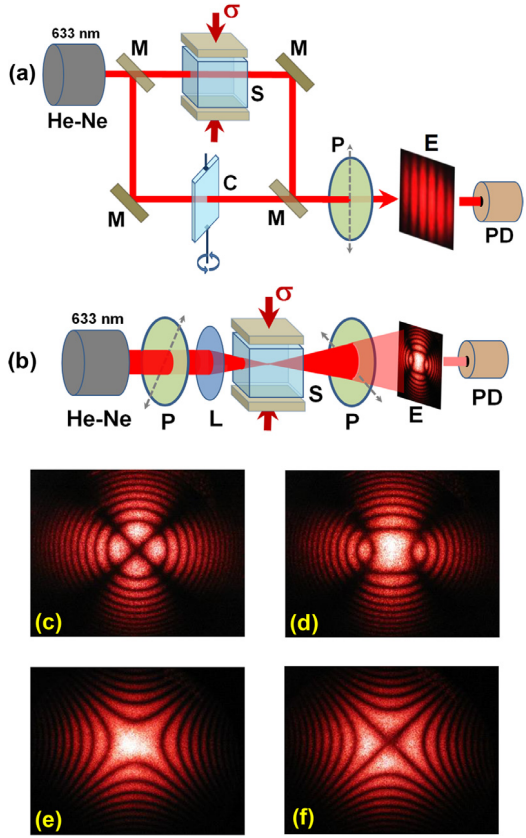

**Fig. 2.** Experimental setups for piezooptic measurements combining Mach-Zehnder (a) and conoscopic (b) interferometry. *M* - mirrors, *C* - optical pathlength compensator, *P* – polarizers, *L* – lense, *E* – screens, *PD* – photodetectors. Sample *S* is subjected to uniaxial compression $\sigma$. Sections (c) and (d) demonstrate conoscopic images captured at $\sigma_2 = 0$ and $\sigma_2 = 230$ kgf/cm², respectively, for transverse piezooptic coupling geometry ($k = 3$, $m = 2$). Sections (e) and (f) show conoscopic images captured at $\sigma_2 = 0$ and $\sigma_2 = 180$ kgf/cm², respectively, for transverse piezooptic coupling geometry ($k = 1$, $m = 2$). Applied half-wave stress changes the light intensity, measured by the photodetector in the central part of the conoscopic image, from its minimum to maximum value, or vice versa.

One must admit, however, that certain POCs, such as e.g. $\pi_{12}$ or $\pi_{66}$, cannot be measured precisely by interferometric Mach-Zehnder technique. Accordingly, for accurate determination of these POCs we applied a conoscopic interferometry following the methodology described recently by Mytsyk et al. [22,36]. Earlier, Wang et al. [23] have demonstrated an application of similar technique for probing of electro-optic coefficients in strontium calcium barium niobate crystals. Fig. 2(b) shows relevant setup. The coherent He-Ne laser light ($\lambda = 633$ nm) is focused by the short focal lens *L* on the crystal sample *S* being subjected to applied uniaxial stress $\sigma$. The polarizers *P* are crossed and their polarization axes are usually turned by ±45° to the direction of the uniaxial compression $\sigma$. The resulting interference pattern (conoscopic figure) is projected on the screen *E* and captured by the camera. For more precise evaluations a lens system has been used to improve a quality of the conoscopic images projected on the screen, in a similar way suggested recently by Montalto et al. [39,40]. The intensity of the light outgoing through a few millimeter hole, made in the center of the screen, is detected by the photodiode *PD*. The shape of conoscopic figures depends on the sample orientation with respect to propagation direction of divergent light. Figs. 2(c),(d) and Figs. 2(e),(f) show typical conoscopic figures observed when light propagates along the optical axis of LTB crystal and perpendicular to it, respectively. The uniaxial stress applied to the sample modifies the pattern, as evidenced by panels (d) and (f) of this figure, changing the light intensity in the central part of the conoscopic figures. Accordingly, the retardation half-wave stress value, $\sigma_m^{*\lambda/2}$, extracted in such measurements, is defined in a similar way as in the interferometric measurements, i.e. as absolute difference $|\sigma_{max} - \sigma_{min}|$, where $\sigma_{max}$ and $\sigma_{min}$ are the stress magnitudes corresponding to neighbor maximum and minimum values of a stress dependent light intensity $I(\sigma)$ measured in the central part of the conoscopic figure. Relevant interference patterns are presented in Fig. 2(d) and 2(e), as being captured at $\sigma_{max}$, whereas Fig. 2(c) and 2(f) demonstrate the interference patterns shot by camera at $\sigma_{min}$.

Certain challenges in piezooptic measurements arise often regarding the off-diagonal POCs $\pi_{61}$, $\pi_{16}$ or $\pi_{45}$. Usually they are determined by measuring a rotation of the optical indicatrix under mechanical stress action as described, particularly, in Refs. [22,36]. Fortunately, due to symmetry reasons these tensor components equal zero for 4*mm* point-group of LTB crystals. Accordingly, the conoscopic interferometry have been applied in the present work basically for the determination of the difference of piezooptic coefficients, $\pi_{km}^o$, characterizing stress-induced optical retardation, where the indices *k* and *m* specify directions of light propagation and uniaxial compression, respectively. Combining then the conoscopic and Mach-Zehnder interferometry one may considerably improve an accuracy of piezooptic measurements what will be discussed in details in Section 4.

## 3. Basic relationships

The piezo-induced optical pathlength, measured by interferometry techniques, consists of two contributions that originate from (i) direct piezooptic effect, caused by stress-induced changes of the refractive indices and (ii) indirect piezooptic effect resulting from a change of the sample thickness in the direction of light propagation induced by a transverse uniaxial compression. Precise determination of POCs requires thereby a consideration of both effects what in the case of principal POCs, $\pi_{im}$ (*i*, *m* = 1, 2, 3), measured by the interferometric half-wave method, leads to simple known relation [37,41,42]:

$$\pi_{im} = -\frac{\lambda}{\sigma_{im}^{\lambda/2} n_i^3 d_k} + \frac{2S_{km}}{n_i^3}(n_i - 1) = -\frac{\lambda}{\sigma_{im}^o n_i^3} + \frac{2S_{km}}{n_i^3}(n_i - 1), \qquad (1)$$

where $n_i$ is the refractive index, $d_k$ is the sample thickness along the light propagation, $S_{km}$ is the elastic compliance tensor constant and $\sigma_{im}^o = \sigma_{im}^{\lambda/2} d_k$ is the thickness normalized half-wave stress, i.e. the value independent on sample dimensions commonly used for catheterization of piezooptic materials. The indices *k*, *i* and *m* specify directions of light propagation, light polarization and uniaxial compression in Cartesian crystallophysic coordinate system, respectively. One should notice that for the crystals of tetragonal symmetry the crystallophysic and crystallographic systems coincide. Considering, for instance, the light polarized parallel to the *X*-axis (*i* = 1), which propagates along the crystal optical axis (*k* = 3, crystallographic *Z*-axis), and uniaxial compressions $\sigma_1$ or $\sigma_2$, applied along the *X*-axis (*m* = 1) or *Y*-axis (*m* = 2), respectively, one obtains:

$$\pi_{11} = -\frac{\lambda}{n_1^3 \sigma_{11}^o} + \frac{2S_{13}}{n_1^3}(n_1 - 1); \quad \pi_{12} = -\frac{\lambda}{n_1^3 \sigma_{12}^o} + \frac{2S_{13}}{n_1^3}(n_1 - 1). \qquad (2)$$

Here we have taken into account that the fourth rank tensor of the elastic compliance is symmetric with respect to the pairs of indices ($S_{km} = S_{mk}$), thereby $S_{31} = S_{13}$ and $S_{32} = S_{23}$. Tetragonal symmetry 4*mm*, in addition, reduces a number of independent tensor components, particularly, $S_{13} = S_{23}$. Similar expressions may be easily derived for all the other principal POCs, $\pi_{im}$ (*k*, *i*, *m* = 1, 2, 3), related, particularly, with the piezooptic measurements performed on the principal-cut-sample [Fig. 1(b)]. Relevant analysis for the *X*/45°- [Fig. 1(c)] or *Z*/45°- [Fig. 1(d)] cuts, on the other hand, results in more complicated

**Table 1**

Basic relationships for the determination of POCs or their sums combinations performing the piezooptic interferometric measurements on $X/45°$- and $Z/45°$- crystal cuts of tetragonal symmetry ($4mm$). Indices $k$, $i$, $m$ specify the experimental geometry, i.e. the directions light propagation, light polarization and uniaxial compression, respectively.

| Experimental geometry | Relationships $X/45°$-cut | Equation number |
|---|---|---|
| $m=4(\bar{4})k=\bar{4}\,(4)i=4(\bar{4})$ | $\pi_{11}+\pi_{13}+\pi_{31}+\pi_{33}+2\pi_{44}=-\frac{4\lambda}{n_4^3\sigma_{44(\bar{4}\bar{4})}^o}$ $+2(S_{11}+2S_{13}+S_{33}-S_{44})\frac{(n_4-1)}{n_4^3}$ | (T.1) |
| $m=1k=4(\bar{4})i=1$ | $\pi_{11}=-\frac{\lambda}{n_1^3\sigma_{11}^o|_{k=4(\bar{4})}}+(S_{12}+S_{13})\frac{n_1-1}{n_1^3}$ | (T.2) |
| $m=4(\bar{4})k=\bar{4}\,(4)i=1$ | $\pi_{12}+\pi_{13}=-\frac{2\lambda}{n_4^3\sigma_{14(1\bar{1})}^o}+(S_{11}+2S_{13}+S_{33}-S_{44})\frac{n_1-1}{n_1^3}$ | (T.3) |
| $m=1k=\bar{4}\,(4)i=4(\bar{4})$ | $\pi_{12}+\pi_{31}=-\frac{2\lambda}{n_4^3\sigma_{41(\bar{4}1)}^o}+2(S_{12}+S_{13})\frac{n_1-1}{n_4^3}$ | (T.4) |
| $m=4(\bar{4})k=1i=2$ | $\pi_{11}+\pi_{13}=-\frac{2\lambda}{n_1^3\sigma_{24(2\bar{4})}^o}+2(S_{12}+S_{13})\frac{n_1-1}{n_1^3}$ | (T.5) |
| $m=4(\bar{4})k=1i=3$ | $\pi_{31}+\pi_{33}=-\frac{2\lambda}{n_3^3\sigma_{34(3\bar{4})}^o}+2(S_{12}+S_{13})\frac{n_1-1}{n_3^3}$ | (T.6) |
| $Z/45°$-cut | | |
| $m=3k=6(\bar{6})i=3$ | $\pi_{33}=-\frac{\lambda}{n_3^3\sigma_{33}^o|_{k=6(\bar{6})}}+2S_{13}\frac{n_3-1}{n_3^3}$ | (T.7) |
| $m=3k=\bar{6}\,(6)i=6(\bar{6})$ | $\pi_{13}=-\frac{\lambda}{n_1^3\sigma_{63(\bar{6}3)}^o}+2S_{13}\frac{n_1-1}{n_1^3}$ | (T.8) |
| $m=6(\bar{6})k=\bar{6}\,(6)i=6(\bar{6})$ | $\pi_{11}+\pi_{12}+\pi_{66}=-\frac{2\lambda}{n_1^3\sigma_{66(\bar{6}\bar{6})}^o}+(2S_{11}+2S_{12}-S_{66})\frac{n_1-1}{n_1^3}$ | (T.9) |
| $m=6(\bar{6})k=\bar{6}\,(6)i=3$ | $\pi_{31}=-\frac{\lambda}{n_3^3\sigma_{36(3\bar{6})}^o}+\frac{1}{2}(2S_{11}+2S_{12}-S_{66})\frac{n_3-1}{n_3^3}$ | (T.10) |

expressions. For the tetragonal point group $4/m$ they have been already derived in Ref. [43]. The point group $4mm$, however, is characterized by a smaller number of independent non-zero piezooptic constants, particularly $\pi_{61}=\pi_{16}=\pi_{45}=0$. Accordingly, it reduces a number of equations likewise the required numbers of the samples and measurements which has to be done in order to determine the full set of POCs matrix, see Table 1. Notice, that equations presented there are suitable for the determination of individual POCs ($\pi_{11}$, $\pi_{33}$, $\pi_{13}$, $\pi_{31}$) as well as possible sums combinations, as e.g. $\pi_{12}+\pi_{13}$, $\pi_{12}+\pi_{31}$, $\pi_{11}+\pi_{12}+\pi_{66}$ and several others. The individual POCs and/or their sums being determined by independent measurements on different samples may serve here for verification of the piezooptic measurements, i.e. as the most reliable method for accuracy validation. More detailed discussion on this issue will be given in the next section.

### 4. Experimental results and analysis

Determination of POCs by interference technique is based on half-wave stress measurements performed on the same or different samples for different geometries of piezooptic coupling defined by directions of light propagation, light polarization and uniaxial compression. Table 2 lists thickness normalized half-wave stress magnitudes, $\sigma_{im}^o$, of LTB crystals, determined from interferometric measurements, and relevant magnitudes of POCs consequently calculated by means of equations set (T.1)-(T.10) given in Table 1. In our calculations the effective refractive indices $n_1=n_2=n_6=n_o$ and $n_3=n_e$, where $n_o=1.6088$ and $n_e=1.5520$ are the magnitudes of the ordinary and extraordinary refractive indices of LTB crystals, respectively, taken from Ref. [44]. The effective refractive indices along the diagonal crystallographic directions, 4 or $\bar{4}$ [see Fig. 1(c)] have been determined as:

$$n_4=n_{\bar{4}}=\frac{\sqrt{2}}{\sqrt{a_2+a_3}}=\frac{\sqrt{2}}{\sqrt{\frac{1}{n_2^2}+\frac{1}{n_3^2}}}=\sqrt{2}n_2n_3/(n_2^2+n_3^2)^{1/2}=1.5796, \quad (3)$$

where $a_2=n_2^{-2}$ and $a_3=n_3^{-2}$ are the optical polarization constants. A crucial issue in such calculations was a choice of proper values of the elastic compliance constants, $S_{km}$, calculated by inversion of the elastic constant matrix ($S=C^{-1}$), taking into account discrepancy between values for certain elastic constants $C_{ij}$ reported in the literature by different authors [30,45,46], see Table 3. Our analysis shows, that the best convergence of POCs values determined for different samples and/or

**Table 2**

The thickness normalized half-wave stress magnitudes, $\sigma_{im}^o$, as obtained from piezooptic interferometric measurements, and calculated POCs $\pi_{im}$ of LTB crystals, $\lambda=633$ nm, $T=20$ °C.

| No | Experimental geometry | | | $\sigma_{im}^o$, kgf/cm | $\pi_{im}$, Br |
|---|---|---|---|---|---|
| | $m$ | $k$ | $i$ | | |
| Principal-cut-sample | | | | | |
| 1. | 1 | 2 | 1 | $\sigma_{11}^o=-210$ | $\pi_{11}=-0.38\pm0.08$ |
| 2. | 1 | 2 | 3 | $\sigma_{31}^o=195$ | $\pi_{31}=1.25\pm0.10$ |
| 3. | 1 | 3 | 1 | $\sigma_{11}^o=114$ | $\pi_{11}=-0.37\pm0.22$ |
| 4. | 1 | 3 | 2 | $\sigma_{21}^o=71$ | $\pi_{21}=0.45\pm0.28$ |
| 5. | 2 | 1 | 2 | $\sigma_{22}^o=-215$ | $\pi_{22}=-0.36\pm0.08$ |
| 6. | 2 | 1 | 3 | $\sigma_{32}^o=205$ | $\pi_{32}=1.21\pm0.09$ |
| 7. | 2 | 3 | 2 | $\sigma_{22}^o=120$ | $\pi_{22}=-0.44\pm0.22$ |
| 8. | 2 | 3 | 1 | $\sigma_{12}^o=75$ | $\pi_{12}=0.34\pm0.27$ |
| 9. | 3 | 1 | 3 | $\sigma_{33}^o=57$ | $\pi_{33}=1.28\pm0.35$ |
| 10. | 3 | 1 | 2 | $\sigma_{23}^o=28$ | $\pi_{23}=3.81\pm0.58$ |
| 11. | 3 | 2 | 3 | $\sigma_{33}^o=56$ | $\pi_{33}=1.34\pm0.35$ |
| 12. | 3 | 2 | 1 | $\sigma_{13}^o=30$ | $\pi_{13}=3.44\pm0.54$ |
| $X/45°$-cut | | | | | |
| 13. | 1 | 4 | 1 | $\sigma_{11}^o=580$ | $\pi_{11}=-0.42\pm0.09$ |
| 14. | 1 | 4 | $\bar{4}$ | $\sigma_{41}^o=110$ | $\pi_{12}+\pi_{31}=1.60\pm0.35$ |
| 15. | 1 | $\bar{4}$ | 1 | $\sigma_{11}^o=600$ | $\pi_{11}=-0.43\pm0.09$ |
| 16. | 1 | $\bar{4}$ | 4 | $\sigma_{41}^o=115$ | $\pi_{12}+\pi_{31}=1.47\pm0.34$ |
| 17. | 4 | $\bar{4}$ | 4 | $\sigma_{44}^o=275$ | $\pi_{44}=-1.00\pm0.61$ |
| 18. | 4 | $\bar{4}$ | 1 | $\sigma_{14}^o=88$ | $\pi_{12}+\pi_{13}=4.18\pm0.60$ |
| 19. | 4 | 1 | 2 | $\sigma_{24}^o=69$ | $\pi_{11}+\pi_{13}=3.12\pm0.48$ |
| 20. | 4 | 1 | 3 | $\sigma_{34}^o=77$ | $\pi_{31}+\pi_{33}=3.10\pm0.48$ |
| $Z/45°$-cut | | | | | |
| 21. | 3 | 6 | 3 | $\sigma_{33}^o=59$ | $\pi_{33}=1.18\pm0.34$ |
| 22. | 3 | 6 | $\bar{6}$ | $\sigma_{63}^o=29$ | $\pi_{13}=3.62\pm0.55$ |
| 23. | 3 | $\bar{6}$ | 3 | $\sigma_{33}^o=57$ | $\pi_{33}=1.28\pm0.35$ |
| 24. | 3 | $\bar{6}$ | 6 | $\sigma_{63}^o=29.5$ | $\pi_{13}=3.52\pm0.55$ |
| 25. | 6 | $\bar{6}$ | 6 | $\sigma_{66}^o=-700$ | $\pi_{66}=-0.69\pm0.39$ |
| 26. | 6 | $\bar{6}$ | 3 | $\sigma_{36}^o=135$ | $\pi_{31}=1.18\pm0.18$ |

geometries of piezooptic coupling provides the elastic compliance values reported by Shiosaki et al. [47], see row 8 in Table 3. Here we imply the sample geometries giving directly POCs magnitudes, as derived from the measurements of the principal-cut-sample [Fig. 1(b)], or their sums combinations, as obtained from the measurements of $X/45°$- [Fig. 1(c)] or $Z/45°$- [Fig. 1(d)] sample cuts. In turn, for the calculation of photoelastic constants, $p_{ik}$, we have used the elastic constants, $C_{mk}$ (see Table 3, row 4), obtained by inversion of elastic compliance matrix ($C=S^{-1}$) using $S_{km}$ values from Ref. [47].

**Table 3**
Elastic constants $C_{mk}$ ($\times 10^{10}$ N/m$^2$) and elastic compliances $S_{km}$ ($\times 10^{-12}$ m$^2$/N) of LTB crystals.

| $C_{mk}$ | $C_{11}$ | $C_{33}$ | $C_{12}$ | $C_{13}$ | $C_{44}$ | $C_{66}$ | |
|---|---|---|---|---|---|---|---|
| 1 | 13.29 | 6.73 | 0.57 | 4.05 | 5.58 | 4.91 | Ref. [43] |
| 2 | 13.10 | 6.14 | 0.95 | 3.85 | 5.50 | 4.79 | Ref. [30] |
| 3 | 13.53 | 5.48 | 0.11 | 3.19 | 5.74 | 4.74 | Ref. [44] |
| 4 | **13.55** | **5.68** | **0.36** | **3.35** | **5.85** | **4.67** | Calculated using values of row 8 |

| $S_{km}$ | $S_{11}$ | $S_{33}$ | $S_{12}$ | $S_{13}$ | $S_{44}$ | $S_{66}$ | |
|---|---|---|---|---|---|---|---|
| 5 | 9.50 | 22.92 | 1.63 | −6.70 | 17.92 | 20.37 | Calculated using values of row 1 |
| 6 | 9.54 | 24.81 | 1.31 | −6.80 | 18.18 | 20.88 | Calculated using values of row 2 |
| 7 | 8.76 | 25.08 | 1.31 | −5.86 | 17.42 | 21.10 | Calculated using values of row 3 |
| 8 | **8.81** | **24.60** | **1.23** | **−5.92** | **17.10** | **21.40** | Ref. [45] |

**Table 4**
POCs $\pi_{im}$ [in Brewsters (Br), 1 Br = 10$^{-12}$ m$^2$/N] and photoelastic constants, $p_{ik}$, of LTB crystals.

| $\pi_{11}$ | $^*\pi_{12}$ | $\pi_{13}$ | $\pi_{31}$ | $\pi_{33}$ | $\pi_{44}$ | $^*\pi_{66}$ |
|---|---|---|---|---|---|---|
| − 0.40 ± 0.09 | 0.46 ± 0.11 | 3.60 ± 0.56 | 1.23 ± 0.10 | 1.27 ± 0.35 | − 1.00 ± 0.61 | − 0.74 ± 0.05 |

| $p_{11}$ | $p_{12}$ | $p_{13}$ | $p_{31}$ | $p_{33}$ | $p_{44}$ | $p_{66}$ |
|---|---|---|---|---|---|---|
| 0.068 ± 0.026 | 0.181 ± 0.028 | 0.206 ± 0.038 | 0.214 ± 0.025 | 0.155 ± 0.022 | − 0.059 ± 0.036 | − 0.035 ± 0.004 |

*POCs determined by the conoscopic interferometry method.

Several comments may be useful in understanding of the obtained results.

(i) Following Table 2 most POCs $\pi_{im}$ ($i, m = 1, 2, 3$) are determined applying different geometries of piezooptic coupling. For example, taking into account that $\pi_{11} = \pi_{22}$ POC $\pi_{11}$ has been determined for six different such geometries using both the principal-cut-sample (see rows 1, 3, 5 and 7) and the $X/45°$-cut (see rows 13, 15). The magnitudes provided in rows 1, 5, 13, 15 are characterized by considerably smallest errors thus the resulting $\pi_{11}$-value given in Table 4 represents their average. For comparison, POCs values $\pi_{11}$ ($\pi_{22}$) determined for piezooptic coupling geometries, as specified by rows 3 or 7 in Table 2, are presented with about three times larger errors. POCs $\pi_{13}$ or $\pi_{23}$ ($\pi_{13} = \pi_{23}$), on the other hand, have been determined in four different coupling geometries performing the measurements on both the principal-cut (rows 10 and 12, Table 2) and the $Z/45°$-cut (rows 22 and 24, Table 2) samples. Since relevant errors are practically the same in all these cases the $\pi_{13}$ magnitude provided in Table 4 represents an average value over these four geometries of piezooptic coupling. In a similar way it was derived also POC $\pi_{33}$ magnitude (rows 9, 11, 21, 23). One should be emphasized that $\pi_{13}$ POC is characterized by a quite large magnitude (3.60 ± 0.56 Br) which appears to be several times larger compared to relevant maximal principal POCs $\pi_{im}$ in a number of efficient photoelastic crystal materials such as e.g. lithium niobate [20,27,37], strontium borate [48], gallium phosphide [41] or calcium wolframate [35,42]. In this context, LTB is comparable by its piezooptic efficiency with lead molybdate [49] or $\beta$-barium borate [9]. Accordingly, it is expected to exhibit considerable photoelasticity and acoustooptic efficiency what will be demonstrated below.

(ii) The values of POCs in Table 2 are provided with relevant errors representing the standard deviations which account for the errors of measured thickness normalized half-wave stresses $\sigma_{im}^o$ and elastic compliances $S_{km}$ reported in the literature [30,45–47]. By analyzing $S_{km}$-values given in Table 3, particularly their mean square deviations, one may accept a relative error for their determination at a level of about 10% in average. In a number of cases the elastic factors in Table 1 represent a superposition of $S_{km}$ constants [see e.g. Eqs. (T1)-(T6)] thus the mean-square deviations of relevant sums have been evaluated in each case. Several diagonal tensor components, on the other hand, such as e.g. POCs $\pi_{44}$ i $\pi_{66}$ are expressed also via principal POCs $\pi_{im}$, see Eqs. (T.1) and (T.9) in Table 1. In such cases errors in determination of the principal POCs have direct influence on the values of other dependent on them POCs rising thereby total errors of their determination.

(iii) Comparing the sums of POCs $\pi_{im}$, as determined on the $X/45°$-cut sample (see rows 14, 16, 18–20 in Table 2), with relevant ones obtained on the principal-cut sample one may admit their practically perfect matching. For instance, the sum $\pi_{12} + \pi_{31}$, being determined on the $X/45°$-cut, equals 1.60 ± 0.35 Br (row 14). It agrees within the measurement error with value of 1.69 ± 0.15 Br obtained as the sum of POCs $\pi_{12}$ and $\pi_{31}$ being determined independently on the principal-cut sample, see relevant values in Table 4. Such good matching appears to be characteristic also for other geometries of piezooptic coupling presented, particularly in Table 2.

(iv) Methodology for determination of POCs $\pi_{12}$ or $\pi_{21}$ ($\pi_{12} = \pi_{21}$) is worth of special attention and more detailed discussion. Mach-Zehnder interferometry indeed gives the $\pi_{21}$-value with a quite large error, $\delta\pi_{12} = \pm 0.28$ Br (see rows 4 i 8 in Table 2), what is about 60% of its absolute magnitude. Accordingly, the conoscopic interferometry (see Fig. 2) may serve here as alternative technique able to improve an accuracy of piezooptic measurements.

In such experiments one measures the light intensity in the central part of the conoscopic pattern. The change of the light intensity from its minimum to maximum value (or vice versa) upon an uniaxial sample compression corresponds to the retardation half-wave stress value $\sigma_{km}^{*\lambda/2}$ being derived from the measurements. In general case one deals with stress-induced optical retardation characterized by so-called piezooptic retardation constant (PORC) $\pi_{km}^o$. In accordance with [50] $\pi_{km}^o$ is determined as:

$$\pi_{km}^o = -\frac{2\delta\Delta}{d_k \sigma_m} = -\frac{\lambda}{d_k \sigma_{km}^{*\lambda/2}} = -\frac{\lambda}{\sigma_{km}^{*o}}, \quad (4)$$

where $\delta\Delta$ is the optical retardation induced by the applied stress $\sigma_m$, $d_k$ is the thickness in the direction of light propagation, $\sigma_{im}^{*o} = \sigma_{im}^{*\lambda/2} d_k$ is the thickness normalized retardation half-wave stress value. In our particular case the light propagates along the crystallographic $Z$-axis ($k = 3$) whereas the uniaxial compres-sion is parallel to the crystallographic Y-axis ($m=2$). For such piezooptic coupling geometry one measures PORC $\pi_{32}^o$ defined as

$$\pi_{32}^o = -\frac{\lambda}{\sigma_{32}^{*o}}, \tag{5}$$

thus with the retardation half-wave stress value $\sigma_{32}^{*o} = 180$ kgf/cm, as derived from our measurements at $\lambda = 633$ nm (uniaxial compression is assigned as negative), one obtains PORC value $\pi_{32}^o = 3.59 \pm 0.25$ Br. Here the relative error has been estimated to be equal to about 7% being typical for polarization-optical measurements. One should emphasize that piezo-induced optical retardation, similarly as piezo-induced pathlength, represents superposition of both direct and indirect contributions. Their separation, generally speaking, appears to be crucially important for a correct characterization of the photoelastic properties in anisotropic materials. Whereas the direct contribution is associated with the piezo-induced birefringence, the indirect one is caused by a stress induced sample deformation, i.e. the sample thickness change in the direction of light propagation. Piezo-induced birefringence is characterized by the piezooptic birefringence constant (POBC) $\pi_{km}^*$, expressed via POCs as $\pi_{km}^* = \pi_{im}n_i^3 - \pi_{jm}n_j^3$, where $n_i$ and $n_j$ are the refractive indices [50,51]. POBC $\pi_{km}^*$ and PORC $\pi_{km}^o$ are related each other by the equation [50]:

$$\pi_{km}^* = \pi_{km}^o + 2\Delta n_k S_{km} \tag{6}$$

where $\Delta n_k$ is the optical birefringence in the direction of light propagation. Along the optical axis ($\mathbf{k} \parallel \mathbf{Z}$) the optical birefringence equals zero ($\Delta n_k = \Delta n_3 = 0$) thereby $\pi_{32}^* = \pi_{32}^o$. For the tetragonal symmetry of LTB crystals, $\pi_{22} = \pi_{11}$ and $n_2 = n_1$, thus POBC $\pi_{32}^*$ takes a form

$$\pi_{32}^* = \pi_{12}n_1^3 - \pi_{22}n_2^3 = \pi_{12}n_1^3 - \pi_{11}n_1^3; \tag{7}$$

leading finally to a simple expression for POC $\pi_{12}$:

$$\pi_{12} = \pi_{32}^*/n_1^3 + \pi_{11}. \tag{8}$$

By substituting the magnitudes of piezooptic coefficients into Eq. (8), $\pi_{11} = -0.40 \pm 0.09$ Br and $\pi_{32}^* = \pi_{32}^o = 3.59 \pm 0.25$ Br (see Table 4), as determined by Mach-Zehnder and conoscopic interferometric techniques, respectively, one obtains $\pi_{12} = 0.46$ Br with a substantially smaller error magnitude, $\pm 0.11$ Br. Finally, this value is given in Table 4. The combination of both methods, whenever it is applicable, brings evident benefits consisting, particularly, in considerably improved accuracy of piezooptic measurements.

(v) The efficiency of the conoscopic interferometry was also examined for other POCs, particularly diagonal tensor components $\pi_{44}$ and $\pi_{66}$. As for $\pi_{44}$-component it does not result in a better accuracy compared to Mach-Zehnder interferometry. The reason for this is evident: within the conoscopic interferometry method POC $\pi_{44}$ in the crystals of 4mm-symmetry is expressed as a combination of several principal POCs thus the error value, $\delta\pi_{44}$, represents a superposition of relevant error contributions which in LTB crystals is apparently large. The magnitude of $\pi_{66}$-constant, in contrast, may be determined quite accurately using a simple relation between POC $\pi_{66}$ and PORC $\pi_{36}^o = \pi_{36}^*$ [see Eq. (6)] which reads as [50]:

$$\pi_{66} = \pi_{36}^o/n_1^3. \tag{9}$$

The measurements have been performed on Z/45°-cut sample [see Fig. 1(c)] placed between the two crossed polarizers. A compressive stress has been applied diagonally to X and Y axes ($m=6$), whereas coherent light beam was directed parallel to the optical axis ($k=3$). For the thickness normalized retardation half-wave stress value $\sigma_{36}^{*o} = 210 \pm 15$ kgf/cm, as derived from the measurements, one gets $\pi_{36}^o = 3.07 \pm 0.21$ Br. By substituting it into Eq. (9) one obtains $\pi_{66} = -0.74 \pm 0.05$ Br. Finally, this magnitude is given in Table 4 since relevant absolute error is about 8 times smaller compared to the one obtained in Mach-Zehnder interferometric measurements.

## 5. Photoelastic constants and acoustooptic efficiency

Tensor of photoelastic constants, $p_{ik}$, appears to be highly important for characterization of both photoelastic and acoustooptical properties of crystal materials, basically from the point of view of their practical applications. Table 4 provides the magnitudes of photoelastic constants calculated using the values of POCs $\pi_{im}$ (Table 4) and elastic constants $C_{mk}$ (Table 3). Relevant tensor relations ($p_{ik} = \pi_{im}C_{mk}$) for 4mm point group symmetry take the forms:

$$\begin{aligned}&p_{11} = \pi_{11}C_{11} + \pi_{12}C_{12} + \pi_{13}C_{13}, p_{31} = \pi_{31}(C_{11} + C_{12}) + \pi_{33}C_{13},\\&p_{12} = \pi_{11}C_{12} + \pi_{12}C_{11} + \pi_{13}C_{13}, p_{33} = 2\pi_{31}C_{13} + \pi_{33}C_{33},\\&p_{13} = (\pi_{11} + \pi_{12})C_{13} + \pi_{13}C_{33}, p_{44} = \pi_{44}C_{44}, p_{66} = \pi_{66}C_{66}.\end{aligned} \tag{10}$$

The errors in determination of photoelastic constants are evaluated as mean square deviations, $\delta(\pi_{im} \cdot C_{mk}) = [(\delta\pi_{im} \cdot C_{mk})^2 + (\pi_{im} \cdot \delta C_{mk})^2]^{1/2}$, applied to each term entering into the set of equations, Eq. (10). Here the errors $\delta\pi_{im}$ are given in Table 4 whereas $\delta C_{mk}$ is accepted to be equal to $0.1 \cdot C_{mk}$.

Acoustooptic efficiency is characterized by the figure of merit, $M_2$, defined as [13]:

$$M_2 = n_i^6 p_{ik}^2/(\rho V_l^3), \tag{11}$$

where $n_i$ is the refractive index, $V_i$ is the sound velocity and $\rho$ is the crystal density. Let's evaluate the acousto-optical figure of merit $M_2$ in LTB crystals for the geometries of acoustooptical coupling characterized by the largest photoelastic constants, particularly $p_{31}$ and $p_{13}$, see Table 4. By substituting corresponding magnitudes into Eq. (11), $p_{31} = 0.214$ (Table 4), $n_3 = 1.552$, $\rho = 2.439 \times 10^3$ kg/m$^3$ [47] and $V_1 = 7340$ m/s [45], one obtains $M_2 = 0.66 \times 10^{-15}$ s$^3$/kg which characterizes the acoustooptic efficiency for the longitudinal ultrasonic wave propagating along crystallographic X-axis and light polarization parallel to the crystallographic Z-axis. By replacing mutually directions of sound propagation and light polarization one may reach much better acousto-optic efficiency despite the fact that corresponding photoelastic constant is somewhat smaller ($p_{13} = 0.206$). Indeed, for the longitudinal ultrasonic wave propagating along the crystallographic Z-axis ($V_3 = 5220$ m/s [45]) and light polarization parallel to the crystallographic X-axis ($n_1 = 1.6088$) one gets $M_2 = 2.12 \cdot 10^{-15}$ s$^3$/kg. Substantially larger acoustooptical efficiency is achieved here due to considerably lower ultrasonic velocity and larger refractivity. For such acoustooptical coupling geometry the figure of merit $M_2$ in LTB crystals is about 5 times larger compared to strontium borate, nowadays the best acoustooptic material in deep-ultraviolet spectral region [48,52].

One must be emphasized, that the best photoelastic and/or acoustooptical efficiency may be achieved by analyzing spatial anisotropies of corresponding effects. Relevant optimization approaches (see e.g. [49,52–55]) require an entire set of the photoelastic tensor constants which for LTB is precisely characterized in the present work. The spatial anisotropy analyses of photoelastic and acoustooptical properties of LTB crystals appear, however, beyond of this study and will be considered elsewhere.

## 6. Conclusion

We have presented here the Mach-Zehnder and conoscopic interferometry techniques aimed to explore photoelastic properties of anisotropic crystal materials. The efficiency of such combined approach is demonstrated on tetragonal LTB crystals. Ambiguities, discrepancies or contradictory values being reported by different authors for these crystals [26,30–32] evidently required a reexamination of their photoelastic properties. Accordingly, special attention in our studies has been paid to accuracy and reliability of piezooptic measurements. For these reasons most POCs and/or their superposition have been determined by performing piezooptic measurements on a set of several sample cuts in different experimental geometries, defined by directions of uniaxial compression, light propagation and polarization. Each POC has been extracted therefore from multiply independent measurements. Particularly, POCs $\pi_{12}$ ($\pi_{21}$) and $\pi_{31}$ ($\pi_{32}$) have been derived from two, whereas POCs $\pi_{11}$ ($\pi_{22}$), $\pi_{13}$ ($\pi_{23}$) and $\pi_{33}$ even from four different experimental geometries. POCs $\pi_{12}$ and $\pi_{66}$, on the other hand, have been measured by the conoscopic interferometry method. Reliability of our piezooptic measurements is also proved by the convergence of POCs sums combinations derived from piezooptic measurements on $X/45°$- and $Z/45°$-samples cuts. Comparing actual and previous studies one may admit an acceptable agreement with most POCs values reported in Ref. [26], however, even there one may realize strong discrepancies for several POCs, particularly about 2 times difference for $\pi_{12}$ or $\pi_{66}$ and about 8 times difference for $\pi_{44}$.

Basing on the results of interferometric and conoscopic piezooptic measurements we have determined a full set of photoelastic tensor constants and evaluated the acoustooptical figure of merit of LTB crystals. For the geometries with strong photoelastic coupling LTB demonstrates quite large acoustooptic performance with figure of merit value, $M_2$, achieving $2.12 \times 10^{-15}$ s$^3$/kg. It is several times larger than that of strontium borate crystals, i.e., the acoustooptical material applicable in the deep ultraviolet spectral region. Taking into account high mechanical strength (>300 кgf/cm$^2$), extremely high laser radiation resistance (40 GW/cm$^2$) [25] and broad window of optical transparency (0.17 ≤ λ ≤ 3.3 μm) [24,25] LTB may be considered as advanced acoustooptic material, especially for deep-ultraviolet applications.

**Declaration of Competing Interest**

The authors declare that they have no known competing financial interests or personal relationships that could have appeared to influence the work reported in this paper.

**Acknowledgement**

The authors acknowledge financial support of the present work from the Ministry of Education and Science of Ukraine (the project Nr 0117U004456 "Modulator"). The presented results are part of a project that has received funding from the European Union Horizon 2020 research and innovation programme under the Marie Sklodowska-Curie grant agreement no. 778156. Support from resources for science in years 2018–2022 granted for the realization of international co-financed project Nr W13/H2020/2018 (Dec. MNiSW 3871/H2020/2018/2) is also acknowledged.

**References**

[1] Narasimhamurty TS. Photoelastic and electrooptic properties of crystals. New York: Plenum Press; 1981.
[2] Auld A., Acoustic fields and waves in solids (Krieger publishing company, malabar, Florida, 1973).
[3] Kemp J. Piezo-optical birefringence modulators: a new use for a long-known effect. J Opt Soc Am 1969;59:950–4.
[4] Wang B, Rockwell RR, Leadbetter A. A polarimeter using two photoelastic modulators. Proc SPIE 2004;5531:367–74.
[5] Xu J, Stroud R. Acousto-Optic devices: principles, design, and applications. New York: Wiley; 1992.
[6] Denz C. Optical neural networks. Wiesbaden: Vieweg+Teubner Verlag; 2017.
[7] Braun KJ, Lytle CR, Kavanaugh JA, Thielen JA, Green AS. A simple, inexpensive photoelastic modulator. Am J Phys 2009;77:13–19.
[8] Koechner W. Q-Switching. Solid-State laser engineering. springer series in optical sciences, 1. New York: Springer; 2006.
[9] Andrushchak AS, Bobitski YV, Kaidan MV, Tybinka BV, Kityk AV, Schranz W. Spatial anisotropy of photoelastic and acoustooptic properties in β-BaB$_2$O$_4$ crystals. Opt Mater 2004;27:619–24.
[10] Andrushchak AS, Chernyhivsky EM, Gotra ZYu, Kaidan MV, Kityk AV, Andrushchak NA, Maksymyuk TA, Mytsyk BG, Schranz W. Spatial anisotropy of the acousto-optical efficiency in lithium niobate crystals. J Appl Phys 2010;108:103118.
[11] Andrushchak AS, Mytsyk BG, Demyanyshyn NM, Kaidan MV, Yurkevych OV, Solskii IM, Kityk AV, Schranz W. Spatial anisotropy of linear electro-optic effect in crystal materials: I. Experimental determination of electro-optic tensor in LiNbO$_3$ by means of interferometric technique. Opt Laser Eng 2009;47:31–8.
[12] Andrushchak AS, Mytsyk BG, Demyanyshyn NM, Kaidan MV, Yurkevych OV, Dumych SS, Kityk AV, Schranz W. Spatial anisotropy of linear electro-optic effect in crystal materials: II. Indicative surfaces as efficient tool for electro-optic coupling optimization in LiNbO$_3$. Opt Laser Eng 2009;47:24–30.
[13] Dixon RW, Cohen MG. A new technique for measuring magnitudes of photoelastic tensors and its application to lithium niobate. Appl Phys Lett 1966;8:205–7.
[14] Hariharan P. Basics of interferometry. USA: Elsevier Academic Press; 2007.
[15] Chevalerias R, Latron Y, Veret C. Methods of interferometry applied to the visualization of flows in wind tunnels. J Opt Soc Am 1957;47:703–6.
[16] Zhang Q, Zhong S, Lin J, Chen W, Luo M, Zhong J, Yu Y, Peng Z, Cheng S. Full-range Fourier-domain optical coherence tomography based on Mach–Zehnder interferometer. Opt Laser Eng 2020;124:105794.
[17] Kok Y, Ireland MJ, Robertson JG, Tuthill PG, Warrington BA, Tango WJ. Low-cost scheme for high-precision dual-wavelength laser metrology. Appl Opt 2013;52:2808–14.
[18] Presti DA, Guarepi V, Videla F, Fasciszewski A, Torchia GA. Intensity modulator fabricated in LiNbO$_3$ by femtosecond laser writing. Opt Laser Eng 2018;111:222–6.
[19] Andrushchak AS, Bobitski YaV, Kaidan MV, Mytsyk BG, Kityk AV, Schranz W. Two–fold interferometric measurements of piezo-optic constants: application to β-BaB$_2$O$_4$ crystals. Opt Laser Technol 2005;37:319–28.
[20] Krupych O, Savaryn V, Vlokh R. Precise determination of full matrix of piezo-optic coefficients with a four-point bending technique: the example of lithium niobate crystals. Appl Opt 2014;53:B1–7.
[21] Mytsyk BG, Demyanyshyn NM, Andrushchak AS, Kost' YaP, Parasyuk OV, Kityk AV. Piezooptical coefficients of La$_3$Ga$_5$SiO$_{14}$ and CaWO$_4$ crystals: a combined optical interferometry and polarization-optical study. Opt Mater 2010;33:26–30.
[22] Mytsyk BG, YaP Kost', Demyanyshyn NM, Gaba VM, Sakharuk OM. Study of piezo-optic effect of calcium tungstate crystals by the conoscopic method. Opt Matter 2015;39:69–73.
[23] Wang A, Gao CHY, Xu JQ, Zhang HJ, ShQ Sun. Conoscopic interferometry for probing electro-optic coefficients of strontium calcium barium niobate crystal. Opt Laser Eng 2011;49:870–3.
[24] Petrov V, Rotermund F, Noack F, Komatsu R, Sugawara T, Uda S. Vacuum ultraviolet application of Li$_2$B$_4$O$_7$ crystals: generation of 100 fs pulses down to 170 nm. J Appl Phys 1998;84:5887–92.
[25] Komatsu R, Sugawara T, Sassa K, Sarukura N, Liu Z, Izumida S. Growth and ultraviolet application of Li$_2$B$_4$O$_7$ crystals: generation of the fourth and fifth harmonics of Nd: Y$_3$Al$_5$O$_{12}$ lasers. Appl Phys Lett 1997;70:3492–4.
[26] Krupych O, Mys O, Kryvyy T, Adamiv V, Burak Y, Vlok R. Photoelastic properties of lithium tetraborate crystals. Appl Opt 2016;55:10457–62.
[27] Shaskolskaya MP. Acoustic crystals. Nauka; 1982.
[28] Allen S. Electro-optic materials and applications. Electronic materials. Miller LS, Mullin JB, editors. New York: Plenum Press; 1991.
[29] Kuznetsov AY, Kruzhalov AV, Ogorodnikov IN, Sobolev AB, Isaenko LI. Electronic structure of lithium tetraborate Li$_2$B$_4$O$_7$crystals. Cluster calculations and x-ray photoelectron spectroscopy. Phys Solid State 1999;41:48–50.
[30] Martynyuk-Lototska I, Mys O, Adamiv V, Ya Burak, Elastical Vlokh R. piezooptical and acoustooptical properties of lithium tetra borate crystals. Ukr J Phys Opt 2002;3:264–6.
[31] Martynyuk-Lototska I, Mys O, T Dudok, Adamiv V, Ye Smirnov, Vlokh R. Acousto-optic interaction in α-BaB$_2$O$_4$ and Li$_2$B$_4$O$_7$ crystals. Appl Opt 2008;47:3446–54.
[32] Martynyuk-Lototska I, Dudok T, Mys O, Romanyuk G, Vlokh R. Acoustooptic interaction and photoelastic properties of Li$_2$B$_4$O$_7$ and α-BaB$_2$O$_4$ crystals at the wavelength of 442 nm. Ukr J Phys Opt 2009;10:218–25.
[33] Erba A, Dovesi R. Photoelasticity of crystals from theoretical simulations. Phys Rev B 2013;88:045121.
[34] Mahmoud A, Erba A, KhE El-Kelany, Rérat M, Orlando R. Low-temperature phase of BaTiO$_3$: piezoelectric, dielectric, elastic, and photoelastic properties from ab initio simulations. Phys Rev B 2014;89:045103.
[35] Erba A, Ruggiero MT, Korter TM, Dovesi R. Piezo-optic tensor of crystals from quantum-mechanical calculations. J Chem Phys 2015;143:144504.
[36] Mytsyk BG, Demyanyshyn NM, Solskii IM, Sakharuk OM. Piezo- and elasto-optic coefficients for calcium tungstate crystals. Appl Opt 2016;55:9160–5.
[37] Mytsyk BG, Andrushchak AS, Demyanyshyn NM, YaP Kost', Kityk AV, Mandracci P, Schranz W. Piezo-optic coefficients of MgO-doped LiNbO$_3$ crystals. Appl Opt 2009;48:1904–11.
[38] Andrushchak AS, Tybinka BV, Ostrovskij IP, Schranz W, Kityk AV. Automated interferometric technique for express analysis of the refractive indices in isotropic and anisotropic optical materials. Opt Laser Eng 2008;46:162–7.
[39] Montalto L, Paone N, Scalise L, Rinaldi D. A photoelastic measurement system for residual stress analysis in scintillating crystals by conoscopic imaging. Rev Sci Instrum 2015;86:063102.
[40] Montalto L, Paone N, Rinaldi D, Scalise L. Inspection of birefringent media by photoelasticity: from diffuse light polariscope to laser conoscopic technique. Opt Eng 2015;54:081210.
[41] Mytsyk BG, Andrushchak AS, YaP Kost'. Static photoelasticity of gallium phosphide crystals. Crystallogr Rep 2012;57:124–30.
[42] Mytsyk BG, YaP Kost', Demyanyshyn NM, Andrushchak AS, Solskii IM. Piezo-optic coefficients of CaWO$_4$ crystals. Crystallogr Rep 2015;60:130–137.
[43] Mytsyk B, Demyanyshyn N, Ya Kost'. Analytical relations describing piezooptic effect in tetragonal crystals. Ukr J Phys Opt 2013;14:101–18.

[44] Sugawara T, Komatsu R, Uda S. Linear and nonlinear optical properties of lithium tetraborate. Solid State Commun 1998;107:233–7.
[45] Kasprowicz D, Kroupa J, Majchrowski A, Michalski E, Drozdowski M, Żmija J. Elastic and nonlinear optical properties of lithium tetraborate. Cryst Res Technol 2003;38:374–8.
[46] Bohatý L, Haussühl S, Liebertz J. Electrooptical coefficients and temperature and pressure derivatives of the elastic constants of tetragonal $Li_2B_4O_7$. Cryst Res Technol 1989;24:1159–63.
[47] Shiosaki T, Adachi M, Kobayashi H, Araki K, Kawabata A. Elastic, piezoelectric, acousto-optic and electro-optic properties of $Li_2B_4O_7$. Jpn J Appl Phys 1985;24:25–7.
[48] MytsykB DemyanyshynN, Martynyuk-Lototska I, Vlokh R. Piezo-optic, photoelastic and acousto-optic properties of $SrB_4O_7$ crystals. Appl Opt 2011;50:3889–95.
[49] Mytsyk B, Erba A, Demyanyshyn N, Sakharuk O. Piezo-optic and elasto-optic effects in lead molibdate crystals. Opt Mater 2016;62:632–8.
[50] Mytsyk B. Methods for the studies of the piezo-optical effect in crystals and the analysis of experimental data. Part I. Methodology for the studies of piezooptical effect. Ukr J Phys Opt 2003;4:1–26.
[51] Sonin AS, Vasylevskaya AS. Electro-optic crystals. Moscow: Atomizdat; 1971.
[52] Demyanyshyn N, Mytsyk B, Sakharuk O. Elasto-optic effect anisotropy in strontium borate crystals. Appl Opt 2014;53:1620–8.
[53] Buryy OA, Andrushchak AS, Kushnir OS, Ubizskii SB, Vynnyk DM, Yurkevych OV, Larchenko AV, Chaban KO, Gotra OZ, Kityk AV. Method of extreme surfaces for optimizing geometry of acousto-optic interactions in crystalline materials: example of $LiNbO_3$ crystals. J Appl Phys 2013;113:083103.
[54] Andrushchak AS, Buryy OA, Demyanyshyn NM, Hotra ZY, Mytsyk BG. Global maxima of the acousto-optic effect in $CaWO_4$ crystals. Acta Phys Pol A 2018;133:928–932.
[55] Mys O, Kostyrko M, Krupych O, Vlokh R. Anisotropy of the acoustooptic figure of merit for $LiNbO_3$ crystals. Isotropic diffraction. Appl Opt 2015;54:8176–86.